\documentclass[aps,showpacs,superscriptaddress,longbibliography]{revtex4-1}
\usepackage{amsfonts}
\usepackage{amssymb}
\usepackage{amsmath}
\usepackage{graphicx,color}
\usepackage{bm}
\usepackage{ulem}
\usepackage{physics}
\usepackage{siunitx}

\begin{document}
%
\title{Breakeven Conditions in Nuclear Fusion via Electron-Free Targets}

\date{\today}

\author{Tadafumi Kishimoto}
\affiliation{Research Center for Nuclear Physics, Osaka University, Ibaraki, Osaka 567-0047, Japan}

\begin{abstract}
Nuclear fusion promises a nearly limitless energy source, but achieving breakeven---where fusion output exceeds input---typically requires extreme plasma conditions and complex confinement systems. Here we propose an alternative approach based on beam--target interactions, introducing a simple energy-based criterion that compares fusion energy generation with energy loss. 
With ideally electron-free targets, stopping power is drastically reduced, enabling conditions in which the breakeven criterion can be satisfied under physically motivated scenarios. This approach offers an alternative conceptual pathway to fusion energy without high-temperature plasma confinement and warrants further investigation.
\end{abstract}

\maketitle

Achieving breakeven in nuclear fusion---where the energy output equals or exceeds the input---remains a central challenge in fusion research. Conventional approaches rely on heating plasma to extreme temperatures, which requires complex confinement systems. 
A recent milestone at the National Ignition Facility (NIF) demonstrated target gain exceeding unity in an indirect-drive implosion~\cite{NIF_PRL_2024_gain}.
In parallel with these advances, our approach explores an alternative pathway by introducing a non-plasma-based breakeven strategy that minimizes energy loss during beam--target interactions.

In contrast, we focus on nuclear reactions involving a beam and a target, and propose a new energy-based criterion that directly compares fusion energy generation with energy loss due to stopping power. This criterion offers a straightforward and practical alternative to the Lawson criterion used in plasma-based fusion systems~\cite{lawson1955,wurzel2022lawson}. We first consider the simplest fusion system consisting only of a beam and a target material. To evaluate breakeven feasibility, we introduce a new index $R(E)$, which compares the energy generated by fusion reactions $E_{\mathrm{gen}}$ to the energy lost due to stopping power $E_{\mathrm{loss}}$ per unit target thickness. This index provides a direct measure of energy generation ratio in beam--target fusion systems and is defined as
\begin{equation}
 R(E) = \dv{E_{\mathrm{gen}}}{x} \,\bigg/\, \dv{E_{\mathrm{loss}}}{x}.
 \label{eq:R}
\end{equation}
Here, $dx$ denotes target thickness expressed as an areal density in \si{\mole\per\centi\meter\squared} (for hydrogen, this coincides numerically with \si{\gram\per\centi\meter\squared}), where the fusion energy generated per unit target thickness $dx$ is
\begin{equation}
 \dv{E_{\mathrm{gen}}}{x} = \rho\,\sigma(E)\,Q,
 \label{eq:gen}
\end{equation}
where $\rho$ is the target density, $\sigma(E)$ is the fusion cross section at beam energy $E$, and $Q$ is the energy released per reaction. The dominant part of the energy loss due to stopping power can be approximated using a simplified form of the Bethe--Bloch formula, which describes the energy loss of fast charged particles due to Coulomb interactions with electrons. The energy-loss term ($E_{\mathrm loss}$) in Eq.~\eqref{eq:R} is evaluated using the Bethe--Bloch stopping-power formula, which we denote as $E_{\mathrm{BB}}$:
\begin{equation}
 \dv{E_{\mathrm{BB}}}{x} \sim K\,\frac{1}{m_e\,\beta^2}\,\ln\!\left(\frac{T_{\max}}{T_{\min}}\right),
 \label{eq:BB}
\end{equation}
where $\beta$ is the beam-particle velocity, $m_e$ is the electron mass and $K$ is a constant that collects target-specific factors, charge information, and units.  $T_{\max}$ and $T_{\min}$ are the maximum and minimum energy transfers in a beam--target collision. 
For simplicity, parameters not essential to the present discussion are absorbed into $K$, while those directly relevant to our discussion are shown explicitly.  The breakeven criterion can be expressed as the requirement that $R(E)>1$, indicating that fusion energy generation exceeds energy loss. 
In this context, the energy loss is compensated by continuous acceleration of the beam, allowing the system to maintain a steady state.
Accordingly, $R(E)>1$ indicates a net energy gain under steady-state operation.

\begin{figure}[t]
 \centering
 \includegraphics[width=0.8\linewidth]{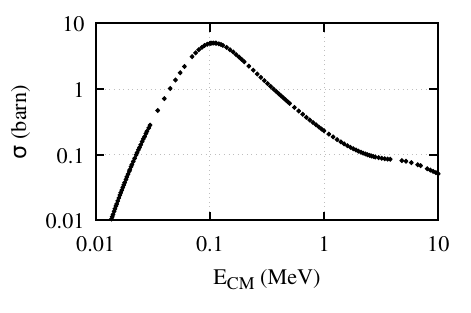}
 \caption{D--T fusion cross section as a function of the center-of-mass energy, taken from the ENDF/B‑VIII.0 evaluated nuclear data library ~\cite{endf2018}.
}
 \label{fig:cs}
\end{figure}

Among various fusion reactions, we focus on the deuterium--tritium (D--T) reaction, which exhibits the highest cross section in the energy range of our interest~\cite{endf2018}. This makes it one of the most promising candidates for achieving breakeven under the proposed criterion. Figure~\ref{fig:cs} shows the D--T reaction cross section as a function of center-of-mass energy.

\begin{figure}[t]
 \centering
 \includegraphics[width=0.8\linewidth]{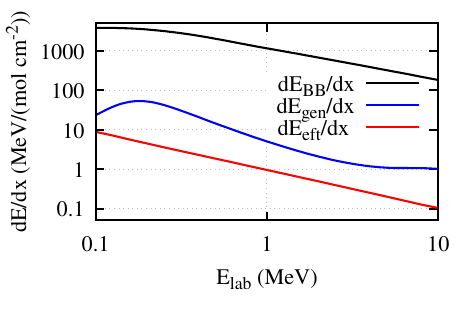}
 \caption{
Fusion energy generation and energy loss per unit target thickness (\si{\mole\per\centi\meter\squared})
as functions of the laboratory-frame beam energy.
The blue curve shows the fusion energy generation rate, $dE_{\mathrm{gen}}/dx$.
The black curve shows the stopping power for conventional electron-containing targets, $dE_{\mathrm{BB}}/dx$, based on NIST 
stopping-power data ~\cite{nist_stopping_power}, which reflect the standard Bethe--Bloch behavior in this energy range.
The red curve shows the stopping power for an electron-free target,
$dE_{\mathrm{eft}}/dx$, evaluated using the cutoff prescription discussed in the text.}
 \label{fig:dedx}
\end{figure}

Using the cross-section data from Fig.~\ref{fig:cs}, we evaluate the energy generation and stopping power as functions of the laboratory-frame beam energy $E_{\mathrm{lab}}$ for a tritium target. Figure~\ref{fig:dedx} compares $\dv*{E_{\mathrm gen}}{x}$ (blue) and $\dv*{E_{\mathrm BB}}{x}$ (black) across $E_{\mathrm{lab}}$. The energy-loss values $\dv*{E_{\mathrm{BB}}}{x}$ are taken from proton stopping-power data in Ref.~\cite{nist_stopping_power} and converted to the corresponding deuteron energy for the same particle velocity. As shown, the energy loss exceeds the generation by approximately two orders of magnitude, yielding $R=\mathcal{O}(10^{-2})$. This two-orders-of-magnitude gap makes breakeven unattainable with conventional beam--target approaches, which explains why high-temperature plasma has been the only viable path. However, this gap is not insurmountable and leaves room for innovative concepts that could mitigate energy losses.

In conventional targets, electrons are present and, as shown in Eq.~\eqref{eq:BB}, they dominate the stopping power due to their low mass. This motivates the concept of electron-free targets, which aim to suppress energy loss by eliminating electrons. The physical origin of energy loss is Coulomb interactions, where the energy transferred per collision scales as
\begin{equation}
 \Delta E \propto \frac{(\Delta p)^2}{2m},
 \label{eq:deltaE}
\end{equation}
where $\Delta p$ is the impulse transferred during the collision and $m$ is the mass of the particle that receives the transferred momentum. Since Coulomb interactions depend solely on the electric charge of the interacting particles, the magnitude of $\Delta p$ is governed by the charge distribution. For the given $\Delta p$, lighter particles receive more energy per collision. This explains why the electron mass appears explicitly in Eq.~\eqref{eq:BB}. In electron-free targets, where only ions are present, the absence of light charged particles significantly reduces stopping power. By replacing electrons with tritium nuclei (mass $m_T$), the energy transferred per collision is reduced by roughly a factor of $m_T/m_e\approx 5500$, 
thereby opening a parameter regime in which fusion reactions become dynamically accessible. 

We next evaluate the logarithmic term in the Bethe--Bloch expression (Eq.~\eqref{eq:BB}). 
In conventional electron-containing materials, the corresponding cutoffs are set by the kinematic upper limit of energy transfer and by atomic binding (ionization) energies, as described by the standard Bethe-–Bloch formulation.  For the electron-free target, the maximum and minimum energy transfers, $T_{\max}$ and $T_{\min}$, are chosen based on the physically relevant range of the impact-parameter ($b$) integral. 
$T_{\max}$ corresponds to the smallest impact parameters ($b_{\mathrm {min}}$) that contribute to the fusion cross section.  Conversely, $T_{\min}$ corresponds to the largest impact parameters ($b_{\mathrm {max}}$), beyond which individual Coulomb interactions no longer lead to appreciable energy transfer.  The processes occurring at $b \le b_{\mathrm {min}}$, which are predominantly due to fusion reactions, effectively remove
projectiles from the beam and therefore act as a sink of particle flux rather than contributing to continuous stopping.
Unlike the standard Bethe--Bloch setting that treats the energy loss of a single projectile,
our formulation must account for a finite-density beam, so the relevant upper cutoff is set on the
beam side by the inter-particle spacing (i.e., by the beam density).
For impact parameters larger than the typical inter-particle spacing, the surrounding ions
form an effectively uniform charge distribution, which exerts no net force on the projectile.
Accordingly, such contributions do not enter the stopping integral, and the upper cutoff $b_{\mathrm {max}}$
can be identified with the mean inter-particle spacing of the beam.  
We therefore adopt a representative beam-particle spacing of \SI{300}{\nano\metre}, set by achievable beam densities rather than by 
the target-ion density. This parameter enters only through the logarithmic term in Eq.~\eqref{eq:BB}, 
so moderate variations in the assumed spacing have little influence on the final result.

While the representative spacing of \SI{300}{\nano\metre} (corresponding to beam densities of
$\sim 10^{12}$–$10^{14}\,\mathrm{cm^{-3}}$) may be optimistic for present low-energy beams, it
illustrates a plausible future regime; even if the beam density is reduced to a more realistic
$10^{9}$–$10^{11}\,\mathrm{cm^{-3}}$, the effective collision count increases by only about one order
of magnitude, so the logarithmic factor varies merely at the $\mathcal{O}(10\%)$ level and our
conclusions remain unchanged.

Figure~\ref{fig:dedx} quantitatively illustrates the consequences of the stopping prescription discussed above, while keeping the fusion cross sections and energy-loss inputs fixed.  It also includes a third curve (red), representing the stopping power for an electron-free target $\dv*{E_{\mathrm{eft}}}{x}$. 
Naively, one might expect the suppression to approach the full electron-to-ion mass ratio of about 5500. In practice, however, the absence of electrons reduces the energy loss by roughly three orders of magnitude (about 1000-2000), due to the logarithmic $\ln(T_{\max}/T_{\min})$ dependence of the 
Bethe-–Bloch stopping power.  
In electron-free targets, $T_{\max}/T_{\min}$ expands to $\sim 10^{9}$–$10^{11}$ from the $\sim10^{3}$–$10^{4}$ range characteristic of conventional targets, since the ionization (binding-energy) threshold that limits $T_{\min}$ in atomic matter is absent and the cutoff is set instead by the inter-particle spacing of the beam.  The logarithmic $\ln(T_{\max}/T_{\min})$ dependence then compresses this vast increase to a factor of only a few, resulting in an effective reduction factor of 1000-2000 rather than the full mass ratio of 5500.

Across the relevant energy range, the ratio 
$\dv*{E_{\mathrm{gen}}}{x}$ to $\dv*{E_{\mathrm{eft}}}{x}$ exceeds unity by a factor of 3--10 under idealized electron-free conditions, suggesting that the breakeven criterion $R(E)>1$ could be met. This represents a significant improvement over conventional targets, where energy loss dominated, while accurate evaluation of residual stopping power, which varies with experimental conditions, remains essential for system design.
Here, we stress that the local breakeven condition $R(E)>1$ does not depend on the
target density (the density cancels in Eq.~(1)), whereas practical power production hinges on achieving sufficient
power density, and thus on raising both target and beam densities.
Ion-only confinement in controlled laboratory settings demonstrates that electron-free ionic configurations are not purely hypothetical.
For intuition, collider‑style pure ion beams in storage rings provide another familiar context in which 
fully stripped ion beams are routinely handled at high phase‑space densities.

Our approach provides an alternative pathway that avoids reliance on high-temperature plasma confinement, which could simplify reactor design and reduce material constraints. Nonetheless, realizing high-density electron-free targets and ensuring beam stability remain significant challenges. While $R(E)$ defined in Eq.~\eqref{eq:R} provides a local criterion under steady-state conditions where energy loss is continuously compensated, the following integration evaluates the total fusion energy generated as the beam slows down from its initial beam energy ($E_B$) to near zero ($E_{\mathrm{fin}}$):
\begin{equation}
E_{\mathrm{gen}} = \int_{E_{\mathrm{fin}}}^{E_B} I(E')\,\rho\,\sigma(E')\,(Q+E')\,
\left(\frac{\mathrm{d}E}{\mathrm{d}x}\Big|_{E'}\right)^{-1}
\,\mathrm{d}E'.
 \label{eq:integral}
\end{equation}
where the stopping power converts the spatial derivative into an energy-based integral, and $I(E)$ denotes the (normalized) beam 
intensity—i.e., the attenuation (survival) factor of the beam flux due to fusion reactions along the path, with $I(E_B)=1$.
Here $E_B$ denotes the initial beam energy at injection, while $E_{\mathrm{lab}}$ is used as the running beam energy in the plots (Figs.~2 and 3).
For clarity, $I(E)$ is given by
\begin{equation}
  I(E)=
  \exp\!\big[-\int_{E}^{E_B}\rho \,\sigma(E')\,
   \left(\frac{\mathrm{d}E}{\mathrm{d}x}\Big|_{E'}\right)^{-1}\,
   \mathrm{d}E'  \ \big]\,
  \label{eq:IofE}
\end{equation}
where $E'$ denotes the integration variable and $E$ is the running beam energy that decreases monotonically from $E_B$ to $E_{\mathrm{fin}}$ along the path.

\begin{figure}[t]
 \centering
 \includegraphics[width=0.8\linewidth]{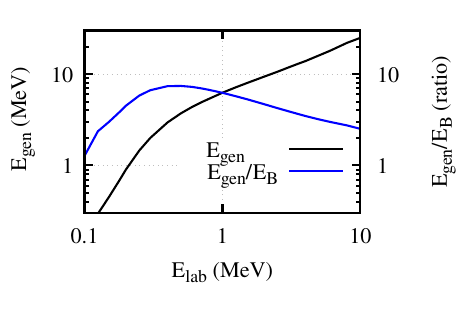}
 \caption{
Integrated fusion energy gain, $E_{\mathrm{gen}}$, accumulated as the beam slows down from its initial energy $E_B$ to near zero, shown as a function of
the initial laboratory-frame beam energy $E_{\mathrm{lab}} \equiv E_B$.
The black curve shows the total fusion energy generated, $E_{\mathrm{gen}}$, in MeV.  The blue curve shows the dimensionless ratio $E_{\mathrm{gen}}/E_B$,
which serves as an integrated breakeven indicator.  The inverse of $E_{\mathrm{gen}}/E_B$ corresponds to the minimum accelerator efficiency
required to avoid net energy loss. 
}
 \label{fig:gain}
\end{figure}

Figure~\ref{fig:gain} illustrates the integrated fusion energy $E_{\mathrm{gen}}$ as the beam slows down, together with the ratio $E_{\mathrm{gen}}/E_B$. The black curve shows $E_{\mathrm{gen}}$ in MeV, while the blue curve shows the dimensionless ratio $E_{\mathrm{gen}}/E_B$, which serves as an integrated breakeven indicator by comparing the total fusion energy generated to the incident beam energy.
For the D–T reaction, $Q=\SI{17.59}{\mega\electronvolt}$, so the theoretical maximum energy gain is $(Q+E_B)/E_B$; for example, at $E_B=\SI{10}{\mega\electronvolt}$, this limit is about $2.76$.  Numerical evaluation shows that the integrated gain reaches about $2.5$
at $E_B=\SI{10}{\mega\electronvolt}$, i.e., close to this theoretical upper bound.
At lower beam energies, the integrated gain can become significantly larger; for example, it reaches about $7$ at $E_B=\SI{0.5}{\mega\electronvolt}$.
Crucially, the integrated gain is governed by how much of the slowing‑down trajectory resides in the energy region where $R(E)>1$. 

Equation~\eqref{eq:R} defines a local (differential) condition for breakeven. By contrast, Eq.~\eqref{eq:integral} evaluates the cumulative fusion energy generated as the beam slows down, from which the integrated breakeven condition is obtained through the ratio $E_{\mathrm{gen}}/E_B$.

Because the beam energy is supplied by an accelerator, its efficiency must be considered when evaluating overall breakeven. We denote the accelerator efficiency by $\varepsilon$, and the effective breakeven index is $R_{\mathrm{eff}}(E)=\varepsilon\,R(E)$, which reflects the practical energy cost of beam generation. In practice, representative efficiencies span a broad range — typically $\varepsilon\!\sim\!0.6$–$0.9$ for electrostatic acceleration and $\varepsilon\!\sim\!0.2$–$0.5$ for RF-based systems — and our conclusions do not rely on the precise choice within these ranges~\cite{handbook_accel}. Importantly, the inverse of this integrated gain, $(E_{\mathrm{gen}}/E_B)^{-1}$, therefore sets the minimum accelerator efficiency required to avoid net energy loss, 
$\varepsilon_{\min}\!\ge\!(E_{\mathrm{gen}}/E_B)^{-1}$. For example, if $E_{\mathrm{gen}}/E_B=2.5$, the accelerator must operate at $\varepsilon\!\ge\!0.4$ to achieve breakeven. This interpretation links the integrated fusion energy gain to practical system design constraints.
At higher beam energies, the impact of finite accelerator efficiency becomes increasingly important. 
As a result, the overall energy balance may favor an intermediate beam-energy range, rather than asymptotically high energies.

This Letter establishes a compact, self-consistent stopping prescription and parameter-transparent bounds that are agnostic to the specific implementation, thereby providing universal design criteria for beam–target fusion, rather than proposing a specific implementation of electron suppression.

\medskip

This work was supported by JSPS KAKENHI Grant Numbers 24H00225, 22H04946. The author also acknowledges support from the Agency for Natural Resources and Energy (Japan).

\bibliography{references}

\end{document}